\pdfoutput=1
\documentclass{ISMA_USD}

\usepackage{booktabs}

\hypersetup{pdftitle  = {Adaptive metamaterial resonators: a discrete and continuous mechanism},
	pdfauthor = {V. Cool, E. Deckers, F. Naets},
	pdfkeywords = {ISMA2026, USD2026, template}}

\title{Adaptive metamaterial resonators: a discrete and \\ continuous mechanism}

\author[1,2]{V. Cool}
\author[2,3]{E. Deckers}
\author[1,2]{F. Naets}

\affil[1]	{KU Leuven, Department of Mechanical Engineering, Division LMSD, \NewLineAffil
			Celestijnenlaan 300, B-3001, Heverlee, Belgium \NewAffil}

\affil[2]	{Flanders Make@KU Leuven, Belgium \NewAffil}

\affil[3]	{ KU Leuven Campus Diepenbeek, Department of Mechanical Engineering, \NewLineAffil 
Wetenschapspark 27, 3590 Diepenbeek, Belgium \NewLineAffil 
			e-mail: \textbf{vanessa.cool@kuleuven.be} }
		
\date{}

\begin{document}



\abstract{Locally resonant metamaterials are an established route to lightweight attenuation of vibration and noise, but their performance is tied to a bandgap fixed at the design stage. In practice, excitation frequencies are rarely constant, as they can vary with operating conditions and loading, causing fixed-frequency resonators to lose effectiveness as conditions drift from their design point. Existing strategies to address this rely on active control, introducing sensors, actuators, and a continuous power supply, or on broadband and multi-resonator designs that trade attenuation depth for bandwidth. A passive alternative, capable of adapting its resonance characteristics directly within the operating environment, would remove this trade-off and open locally resonant metamaterials to a much wider range of real-world, time-varying applications. In this work, we present two adaptive resonator concepts that enable in-situ modification of local resonance characteristics through discrete and continuous adaptation mechanisms.
}


\maketitle


\section{Introduction}

Vibro-acoustic metamaterials are engineered structures designed to manipulate the propagation of elastic and acoustic waves through their internal architecture rather than their constituent materials alone.\
Their dynamic response is typically characterized by frequency ranges, referred to as bandgaps, in which wave propagation is strongly attenuated or completely prohibited.\ Depending on the underlying mechanism, these bandgaps may arise from Bragg scattering or local resonance effects~\cite{Hussein2014,Liu2000}.\ 
Locally resonant metamaterials are particularly attractive for low-frequency attenuation, as they can generate bandgaps in the subwavelength regime.\ 
A key limitation of locally resonant metamaterials is that their bandgaps are typically narrow.\ 
As a result, their attenuation performance can deteriorate significantly when the dominant excitation frequency shifts away from the design frequency.\ 
Such frequency variations are common in practical applications, including rotating machinery, vehicle powertrains, and aerospace structures operating under changing loading conditions.\ 

Several approaches have been explored to extend the operational frequency range of vibro-acoustic metamaterials.\
One strategy aims to broaden the attenuation region itself.\ 
Examples include rainbow metamaterials~\cite{meng2020rainbow}, quasi-zero-stiffness resonators~\cite{cai2022metamaterial}, and topology optimization to fine-tune resonator designs~\cite{cool2026topology}.\ While such approaches enlarge the frequency span over which attenuation occurs, this generally comes at the cost of a reduced peak attenuation level.\
Alternatively, adaptive metamaterials actively tune their resonance frequencies through controllable elements such as piezoelectric patches, shunted circuits, variable-stiffness actuators, magnetic elements, or shape-memory materials~\cite{jian2023adaptive,song2022smoothly}.\ 
These systems can achieve considerable tuning ranges while maintaining strong attenuation across varying operating conditions.\ 
However, they typically require sensing, control electronics, an external power supply, and feedback algorithms, which increase system complexity, cost, and energy consumption.\
A third research path relies on reconfigurable metamaterials, in which bistable elements are exploited to shift the bandgap, here, however, a manual force is currently required to switch from one configuration to the other~\cite{wu2023situ,yao2026reconfigurable}.

This work takes another route and presents two resonator concepts for locally resonant metamaterials that evolve their dynamic properties directly through the mechanical response of the structure, and as such autonomously adapt their bandgap to the excitation frequency.\ 
First, a continuous tuning mechanism is introduced, inspired by the self-tuning sliding-mass concepts encountered in energy harvesting and tuned vibration absorbers~(TVA)~\cite{miller2013experimental,bukhari2021towards}.\ 
Subsequently, a discrete tuning mechanism is presented, based on vibration-induced bistable triggering~\cite{guo2023study}.\ 
For both concepts, the operating principle and mathematical model are introduced, followed by an assessment of their self-adaptation capabilities and the associated advantages and limitations.

The remainder of the paper is organized as follows.\ 
Section~\ref{sec:cont} and Section~\ref{sec:disc} discuss the continuous and discrete mechanisms, respectively.\ 
In each case, the concept and modeling are presented first, after which the results for a particular resonator are discussed.\ 
Finally, Section~\ref{sec:conclusion} summarizes the main findings and conclusions.

\section{Continuous mechanism}
\label{sec:cont}
Inspired by self-tuning sliding-mass~\cite{bukhari2021towards}, the current section investigates the application of this tuning principle to locally resonant metamaterials.\ 
The concept and its mathematical model are first introduced, after which the performance of the proposed mechanism is evaluated for a representative resonator design. 

\subsection{Concept and modeling} 
\label{sec:con_model} 
\begin{figure}
\begin{center}
\includegraphics[width=0.72\linewidth]{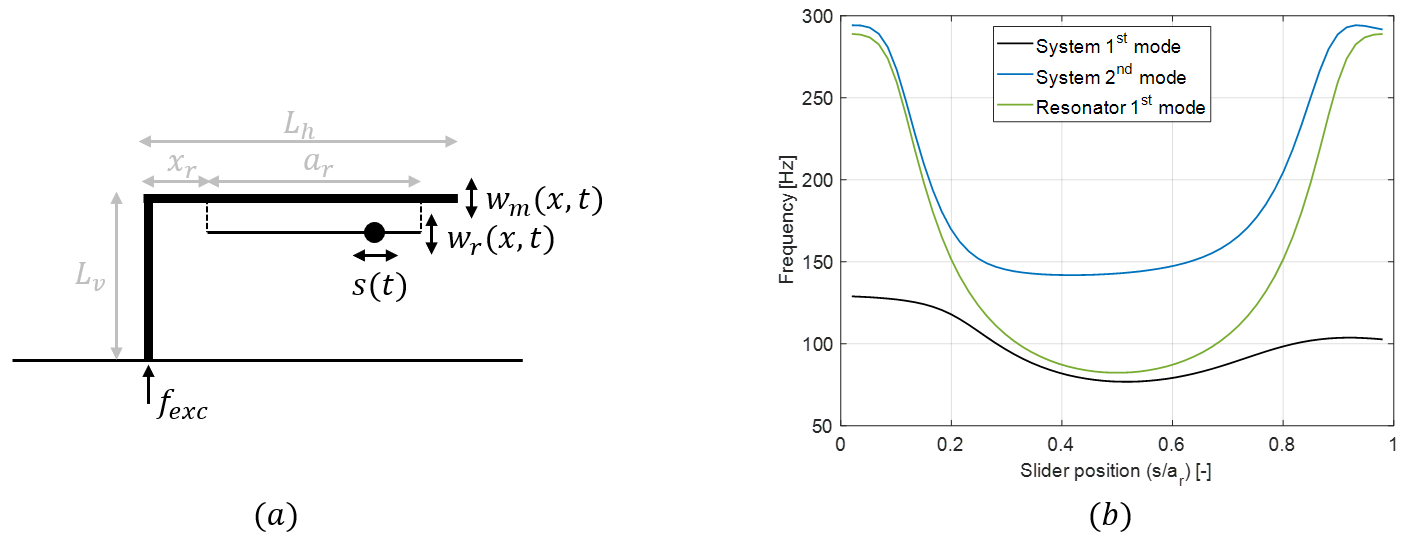}
\caption{a)~Schematic of the L-shaped resonator with sliding mass. b)~Eigenmodes of the L-shaped resonator at different mass position.}
\label{fig:cont1}
\end{center}
\end{figure}

The proposed resonator consists of a slender beam carrying a movable mass $M$, embedded in an L-shaped host structure as shown in Figure~\ref{fig:cont1}.\ 
When subjected to a harmonic base excitation, the slider experiences inertial forces that drive it along the resonator. 
As the slider position changes, the resonator eigenfrequency changes accordingly. 
The mass eventually settles at a location where the resonator frequency matches the excitation frequency, enabling passive self-adaptation.\ 
For the present study, the original straight-cantilever configuration of Bukhari et al.~\cite{bukhari2021towards} is adapted to an L-shaped geometry.\ 
The host structure is excited through a prescribed base displacement $y_0(t)$.
Let $w_m(x,t)$ and $w_r(x,t)$ denote the transverse displacements of the L-beam and slider part relative to the moving base, and let $s(t)$ denote the slider position.
Following~\cite{bukhari2021towards}, the coupled Euler--Bernoulli equations of motion are: \begin{align} 
m_m \ddot{w}_m + E I_m  w_m'''' &= -m_m \ddot{y}_0, \\ 
m_r \ddot{w}_r + E I_r w_r'''' + M\Big[\ddot{w}_r +2\dot{s}\dot{w}'_r +\ddot{s}w'_r +\dot{s}^{\,2}w''_r \Big]_{x=s} &= -m_r\ddot{y}_0 -M\ddot{y}_0\,\delta(x- s), 
\end{align} 
where $m_m$ and $m_r$ are the mass per unit length of the L-beam and resonator, respectively, $EI_m$ and $EI_r$ denote their flexural rigidities, $s$ is the slider position, $\delta(\cdot)$ is the Dirac delta function and $(\dot{  })$, $(')$ represent the derivative with respect to time and space, respectively.\ 
The terms containing $\dot{s}$ and $\ddot{s}$ account for the coupling between the structural vibration and the moving slider.
To efficiently simulate the coupled dynamics, the structural displacements are projected onto a reduced modal basis whose mode shapes are periodically updated as the slider position changes. This yields a nonlinear equation governing the slider motion, 
\begin{equation} 
\ddot s\,(1+\theta^2) = -\theta \left( \ddot w_0 +w_{r,tt} +2\beta\dot s +\kappa\dot s^2 \right), \label{eq:slider_final} 
\end{equation} 
where $w_{r,tt}$ is the local transverse acceleration of the resonator at the slider location, and $\theta$, $\beta$, and $\kappa$ denote the local slope, slope rate, and curvature of the resonator, respectively. 
The moving mass simultaneously exerts a reaction force on the resonator, 
\begin{equation} F_{\mathrm{react}} = -M \left( w_{r,tt} +2\beta\dot s +\theta\ddot s +\kappa\dot s^2 \right), 
\label{eq:freact_final} 
\end{equation} 
which couples the slider dynamics back into the structural response. 
Equations~\eqref{eq:slider_final} and~\eqref{eq:freact_final} constitute the self-tuning mechanism: vibration of the resonator drives the slider motion, while the resulting change in slider position modifies the resonator eigenfrequency until an equilibrium configuration is reached.
The time-domain response is obtained using a two-time-scale integration scheme. At each slow time step $\Delta t_{\rm slow}$ the instantaneous eigenproblem of the slider-loaded structure is solved at the current slider position $s$ to update the modal basis.\ 
Within each slow step, the coupled modal and slider equations are integrated using a fast time step $\Delta t_{\rm fast}$.\ 
Since the slider motion and modal response are mutually dependent through the reaction force $F_{\rm react}$, each time step is resolved through fixed-point iterations before advancing with explicit time integration.\

\subsection{Results} 
\label{sec:res_cont}
A representative resonator design is considered to evaluate the self-tuning principle. The geometric and material parameters are summarized in Table~\ref{tab:params} and are adopted from~\cite{bukhari2021towards}.\ 
Figure~\ref{fig:cont1}b shows the first two eigenfrequencies of the complete resonator together with the eigenfrequency of the clamped sliding part as a function of slider position.\ 
The first eigenfrequency varies from $77$ to $129$~Hz, while the second varies from $142$ to $294$~Hz, in close agreement with the results reported in~\cite{bukhari2021towards}.\ 
Note that the vertical stem of the L-shaped beam is assigned an increased Young's modulus, as a more flexible stem was found to flatten the eigenfrequency variation with slider position, thereby reducing the achievable tuning range.\

\begin{table}
\centering
\caption{Final design parameters for the sliding-mass resonator with continuous tuning mechanism.}
\label{tab:params}
\begin{tabular}{@{}llr@{}}
\toprule
Parameter & Description & Value \\
\midrule
$a_r$ & resonator length & $74.6\,\mathrm{mm}$ \\
$t_r,\,w_r$ & resonator thickness, width & $0.3048\,\mathrm{mm},\ 9.5\,\mathrm{mm}$ \\
$E_r,\,\rho_r$ & resonator modulus, density (steel) & $207\,\mathrm{GPa},\ 7860\,\mathrm{kg/m^3}$ \\
$M$ & slider mass & $7.4\,\mathrm{g}$ \\
$L_h$ & horizontal arm length & $139.7\,\mathrm{mm}$ \\
$x_r$ & resonator start position & $49.4\,\mathrm{mm}$ \\
$t_m,\,w_m$ & host thickness, full width & $3.2\,\mathrm{mm},\ 31.8\,\mathrm{mm}$ \\
$w_{m2}$ & narrow-rail width  & $10.3\,\mathrm{mm}$ \\
$E_m,\,\rho_m$ & host modulus, density & $68.9\,\mathrm{GPa},\ 2700\,\mathrm{kg/m^3}$ \\
$L_v$ & vertical stem length (L-shape only) & $50\,\mathrm{mm}$ \\
-- & stem stiffening factor (E-only) & $10\times$ \\
$\zeta_p$ & modal damping ratio & $0.005$ \\
$c_{\rm slide}$ & slider viscous damping coefficient & $0.15$ \\
\bottomrule
\end{tabular}
\end{table}

\begin{figure}
\begin{center}
\includegraphics[width=\linewidth]{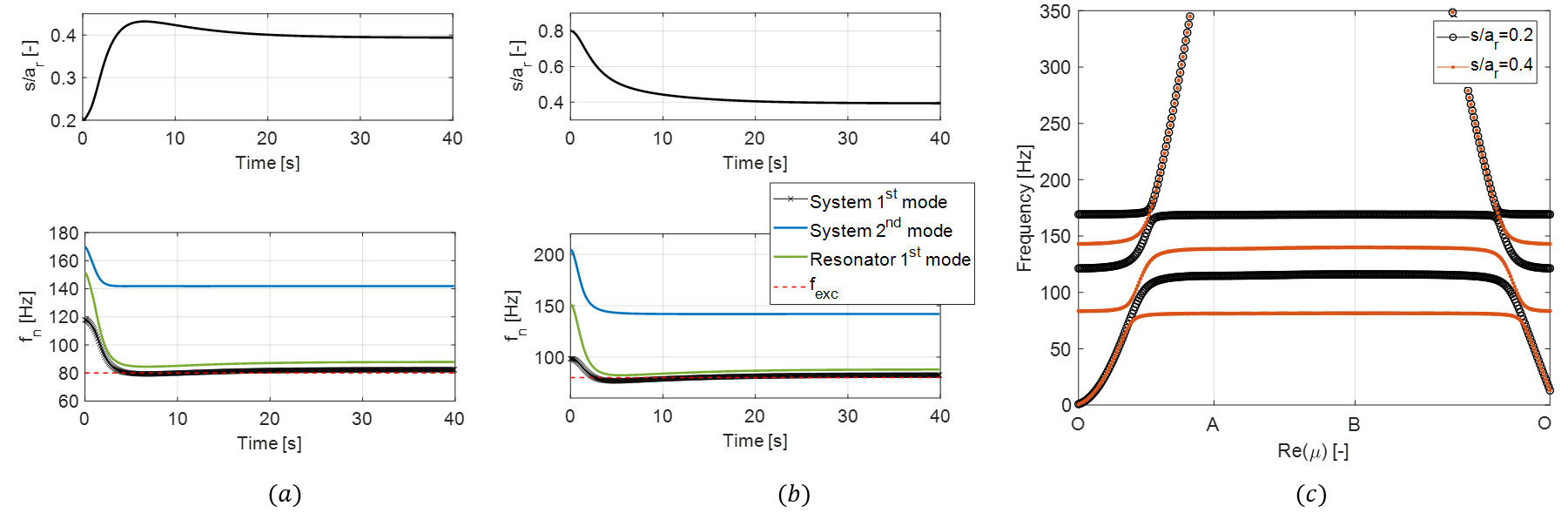}
\caption{Results of the continuous self-adapting resonator. Slider position (top) and corresponding eigenfrequencies (bottom) when exciting with $f_{exc}=80$~Hz starting from a) $s/a_r=0.2$ and b) $s/a_r=0.8$. b) Dispersion curves for the resonator on host structure for slider position  $s/a_r=0.2$ (black curves) showing a bandgap from $115-122$~Hz and  $s/a_r=0.4$ (red curves) showing a bandgap from $81-84$~Hz and $139-143$~Hz.}
\label{fig:cont2}
\end{center}
\end{figure}

Figure~\ref{fig:cont2}a and b show the time-domain response of the resonator under harmonic base excitation at $80$~Hz.\ 
The slider is initialized at $s/a_r=0.2$ and $s/a_r=0.8$ for Figure~\ref{fig:cont2}a and b, respectively, where $a_r$ denotes the total length of the slider.\ 
In both cases, the slider converges towards $s/a_r\approx0.4$, causing the first eigenfrequency of the system to shift from approximately $120$~Hz and $100$~Hz towards the excitation frequency of $80$~Hz.\ 
This demonstrates the self-adapting capability of the proposed concept. However, as noted in the literature on sliding-mass resonators, the slider is generally attracted towards the central region of the beam~\cite{miller2013experimental}.\ 
Consequently, the present mechanism should be viewed as a resonator that self-adjusts to a stable equilibrium position near the beam center, resulting in a local resonance frequency that can adapt within a limited frequency range.\ 
Rather than continuously repositioning over the full beam length to achieve arbitrary resonance frequencies, the slider remains confined to a bounded region, thereby limiting the achievable tuning range~\cite{miller2013experimental}.

Next, the resonator is placed on an aluminum host plate unit cell with dimensions $150\times150\times5$~mm and material properties $E=70$~GPa, $\rho=2700$~kg/m$^3$, and $\nu=0.3$.\ 
These dimensions are selected such that the resonator, which has a width of $139.7$~mm, operates in the subwavelength regime with respect to the host structure.\ 
The resonator is modeled using beam elements and connected to the plate, modeled with solid elements, through four stiff springs.\ 
Figure~\ref{fig:cont2}c shows the dispersion curves for two slider positions, $s/a_r=0.2$ and $s/a_r=0.4$.\ 
The dispersion curves are obtained in MATLAB following the procedure discussed in~\cite{cool2024guide}.\
In both cases, a locally resonant bandgap appears around the first resonance frequency of the resonator, namely near $120$~Hz for $s/a_r=0.2$ and near $80$~Hz for $s/a_r=0.4$.\ 
Note that for the second eigenfrequency, also a bandgap is obtained, for $s/a_r=0.2$ this is a zero-width bandgap, while for $s/a_r=0.4$ a bandgap is present around $140$~Hz.\ 
These results demonstrate that the self-adjustment of the slider towards the excitation frequency results in a corresponding shift of the bandgap, allowing the metamaterial to passively adapt its attenuation range.\ 
It should be noted that the resulting bandgaps are relatively narrow.\ 
This is partly due to the low target frequencies and the deep subwavelength nature of the resonances, but also stems from the resonator design itself.\ 
To enable the sliding mechanism, the resonating element is implemented as a slender clamped-clamped beam embedded within a comparatively stiff L-shaped support.\
Consequently, the resonant motion is largely concentrated in the sliding beam, while the surrounding support exhibits only moderate vibration amplitudes.\
As a result, the force transmitted to the host plate is limited, reducing the achievable bandgap width.\
In conclusion, the sliding-mass resonator demonstrates the ability to passively adapt the frequency range of a locally resonant metamaterial through self-tuning. Nevertheless, the achievable tuning range is constrained by the tendency of the slider to settle near the center of the beam, while the resulting bandgaps remain relatively narrow due to the limited coupling between the resonator and the host structure.

\section{Discrete mechanism}
\label{sec:disc}
Where the previous mechanism was based on a continuous movement of a sliding mass, this section explores a discrete mechanism, inspired by vibration-induced bistable triggering~\cite{guo2023study}.\ After introducing the concept and modeling, the results are discussed for a specific resonator design.

\subsection{Concept \& modeling}
\begin{figure}
\begin{center}
\includegraphics[width=0.5\linewidth]{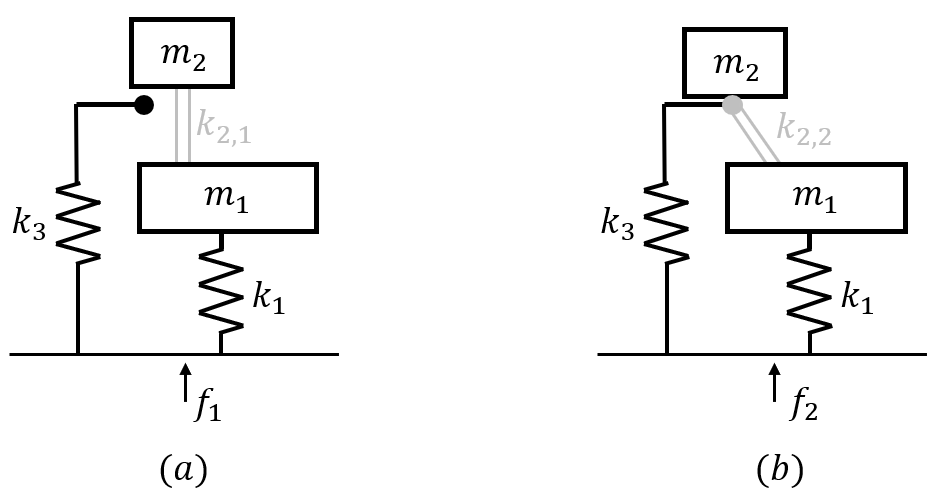}
\caption{Schematic of the two-mass resonator without damping elements: a) pre-snap configuration when excited at $f_1$
  ($k_{2,1}$ coupling, $k_3$ disengaged) and b)
  post-snap configuration triggered by $f_2$ excitation (coupling buckled sideways  at stiffness $k_{2,2}$, ground-path contact $k_3$ engaged).}
\label{fig:discr1}
\end{center}
\end{figure}

The self-tuning resonator with a discrete tuning mechanism is illustrated in Figure~\ref{fig:discr1}.\ 
It consists of a resonator attached to a host structure whose motion is prescribed as $y(t)=Y_0\sin\!\left(2\pi f_{\mathrm{exc}}(t) t\right)$ where $Y_0$ is the excitation amplitude and $f_{\mathrm{exc}}(t)$ is the excitation frequency, which may vary with time.\ 
Figure~\ref{fig:discr1}a shows the initial configuration of the resonator.\ 
The structure consists of two masses ($m_1$ and $m_2$) connected by springs $k_1$ and $k_{2,1}$. Spring $k_1$ is linear, whereas $k_{2,1}$ represents a bilinear element that can buckle laterally under compression.\ 
A third spring, $k_3$, is initially disconnected from the masses.\ 
The system is designed such that the first in-phase resonance occurs at frequency $f_1$, while the second out-of-phase eigenfrequency is located at $f_2$.\ 
When the excitation frequency shifts from $f_1$ to $f_2$, the resulting large relative displacement due to the out-of-phase resonance at that frequency between $m_1$ and $m_2$ can trigger the bistable mechanism, causing the structure to transition to the second configuration shown in Figure~\ref{fig:discr1}b.\ 
This transition is associated with a single contact event that simultaneously performs two functions.\ 
First, the bistable element is latched (e.g., by a magnetic catch or ratchet), replacing its compliant pre-snap stiffness with a significantly stiffer local stiffness $k_{2,2}$.\ 
Second, an independent ground-path spring $k_3$ is engaged.\ 
The values of $k_{2,2}$ and $k_3$ are selected such that the eigenfrequency of the modified configuration coincides with $f_2$.\ 
Consequently, the resonator autonomously retunes itself, allowing its eigenfrequency to match the excitation frequency when operating at $f_2$.\
Note that a separate $k_3$ ground connection is required due to the Cauchy interlacing theorem~\cite{scott1985separation} in multiple degrees-of-freedom systems.\
It should be noted that the proposed topology resembles that of a classical TVA.\ 
However, rather than introducing the secondary mass $m_2$ and spring $k_{2,1}$ to absorb vibrations at a specific frequency, these elements are incorporated to enable the self-tuning functionality of the resonator. 

The system is modeled through the following system of equations:
\begin{align}
  m_1 \ddot{x}_1 &= -c_1(\dot{x}_1 - \dot{y}) - k_1(x_1 - y) + F_c(\delta) + c_2(\dot{x}_2 - \dot{x}_1), \\
  m_2 \ddot{x}_2 &= -F_c(\delta) - c_2(\dot{x}_2 - \dot{x}_1) - F_g ,
\end{align}
with  $y$, $x_1$ and $x_2$, the vertical displacement of the ground, $m_1$ and $m_2$ respectively, $\delta = x_2 - x_1$ and $c_1$ and $c_2$ are the damping coefficient.\ 
$F_c(\delta)$ is the coupling force due to the bistable elements and $F_g$ is the ground-path force through $k_3$.\ 
$F_g$ is zero while the latch is disengaged and
$  F_g = k_3\big((x_2 - y) + \delta_{\mathrm{lock}}\big) + c_3(\dot{x}_2 - \dot{y}) $
once engaged, i.e.\ relaxed at $(x_2-y)=-\delta_{\mathrm{lock}}$.\
In the initial configuration (with the latch disengaged), the internal  coupling force is a piecewise
cubic, softening on the compression side and linear on the extension side:
\begin{equation}
  F_c(\delta) =
  \begin{cases}
    -\hat{k}\, d\,(d - \delta_p)(d - \delta_B), & \delta \le 0,\quad d = -\delta, \\
    k_{\mathrm{ext}}\,\delta, & \delta > 0,
  \end{cases}
  \qquad
  \hat{k} = \dfrac{k_{2,1}}{\delta_p\,\delta_B},
\end{equation}
where $\delta_p$ is the compression (buckling) threshold and $\delta_B$ is the
beam's own second (post-snap) equilibrium. This form is constructed so that
$F_c(\delta)$ has exactly three roots ($\delta=0,\,-\delta_p,\,-\delta_B$)
and reproduces the linear stiffness $k_{2,1}$ near $\delta=0$.\
Once the latch engages, $F_c$ is replaced by a
much stiffer linear spring centered $\delta_{\mathrm{lock}}$: $   F_c(\delta) = k_{2,2}\,(\delta + \delta_{\mathrm{lock}})$.\
The $\delta_{\mathrm{lock}}$-offset in both $F_c$ and $F_g$ after the latch engages guarantees that $\delta=-\delta_{\mathrm{lock}}$ is a genuine
equilibrium of the latched system.\
Further, to avoid the latch flickering on and off every cycle, engagement is
implemented as a persistent two-state hysteretic latch, integrated with
event-based ODE solving.\ 
Both $F_c$ and $F_g$ switch together at the same events: first the latch goes from off to on when $\delta$ decreases through $-\delta_p$, the return on to off latch event is triggered when $\delta$ exceeds $+\delta_{\mathrm{release}}$.
Because $\delta_{\mathrm{release}} > -\delta_p$, a  hysteresis band
separates the engage and release conditions, so ordinary post-snap
oscillation does not cause repeated switching.\

\subsection{Results}
\label{sec:res_discr}
The proposed concept is evaluated for a specific configuration. The host structure consists of an aluminum unit cell with dimensions $30\times30\times5$~mm and material properties $E=70$~GPa, $\rho=2700$~kg/m$^3$, and $\nu=0.3$.\ 
Target frequencies of $200$~Hz and $600$~Hz are considered.\
The total resonator mass is constrained to $50\%$ of the host-structure mass, while the mass ratio $m_1:m_2$ is fixed at $20:1$, since $m_2$ primarily serves to enable the switching mechanism.\ 
The complete set of design parameters is provided in Table~\ref{tab:param_discr}.\ Here, only the rationale behind their selection is discussed. With the masses defined, the stiffnesses $k_1$ and $k_{2,1}$ are chosen such that the first and second eigenfrequencies of the unlatched configuration coincide with $200$~Hz and $600$~Hz, respectively. The cubic spring parameters $\delta_p$, $\delta_B$, and $\delta_{\mathrm{release}}$, together with the damping ratios $\zeta_1=c_1/(2\sqrt{k_1m_1})$ and $\zeta_2=c_2/(2\sqrt{k_{2,1}m_2})$, are carefully tuned to ensure that latching is triggered only when the unlatched configuration is excited at $600$~Hz. For the latched configuration, the stiffnesses $k_3$ and $k_{2,2}$ are selected such that the first in-phase eigenfrequency shifts to $600$~Hz, while the second out-of-phase mode is moved to higher frequencies. With the parameters listed in Table~\ref{tab:param_discr}, the eigenfrequencies of the unlatched configuration are located exactly at $200$~Hz and $600$~Hz. After latching and engagement of the ground-path spring, the first in-phase eigenfrequency shifts to $597$~Hz.

\begin{table}[htbp]
  \centering
  \caption{Final design parameters for the 2 degrees-of-freedom resonator with discrete tuning mechanism.}
  \label{tab:param_discr}
  \begin{tabular}{@{}llr@{}}
    \toprule
    Parameter & Description & Value \\
    \midrule
    $m_1$ & Primary (host-attached) mass & \SI{1.157}{mg} \\
    $m_2$ & Resonator mass & \SI{57.9}{\micro\gram} \\
    $k_1$ & Attachment stiffness & \SI{1930.8}{N/m} \\
    $k_{1,1}$ & Internal coupling, pre-snap & \SI{778.2}{N/m} \\
    $k_{2,2}$ & Internal coupling, post-snap  & \SI{15979}{N/m} \\
    $k_3$ & Ground-path stiffness (engaged on snap) & \SI{141654}{N/m} \\
    $\zeta_1$ & Primary damping ratio & $0.20$ \\
    $\zeta_2$ & Internal coupling damping ratio & $0.003$ \\
    $\zeta_3$ & Ground-path damping ratio & $0.05$ \\
    $Y_0$ & Host excitation amplitude & \SI{1.6}{mm} \\
    $\delta_p$ & Compression (engage) threshold & \SI{2.908}{mm} \\
    $\delta_B$ & Bistable structure's own post-buckling equilibrium & \SI{14.54}{mm} \\
    $\delta_{\mathrm{lock}}$ & Latch-clamped rest position  & \SI{5.00}{mm} \\
    $\delta_{\mathrm{release}}$ & Release threshold & \SI{7.764}{mm} \\
    \bottomrule
  \end{tabular}
\end{table}

\begin{figure}
\begin{center}
\includegraphics[width=0.8\linewidth]{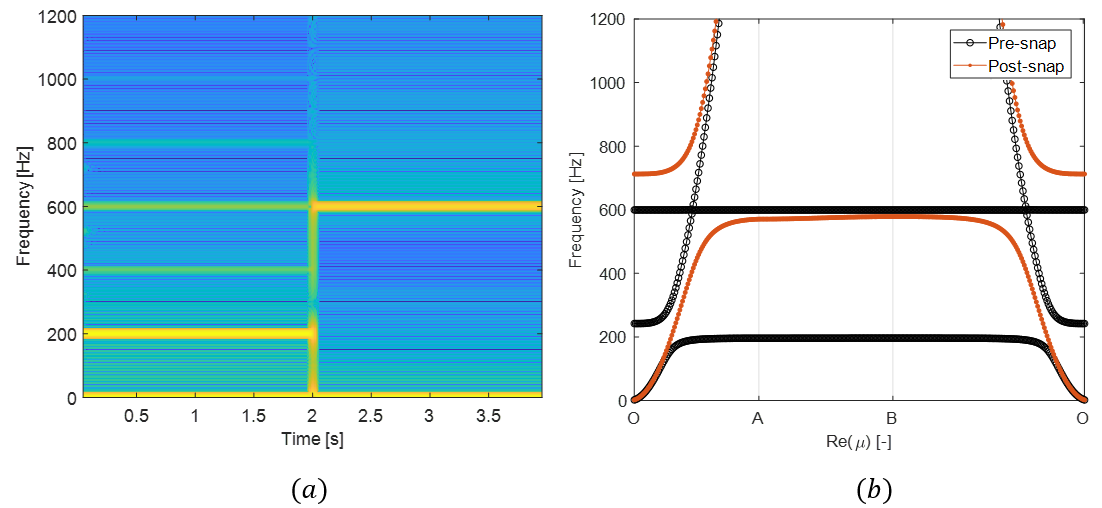}
\caption{Results of the discrete self-adapting unit cell. a) Spectrogram of the $m_2$ response with the colors indicating the power/frequency. b) Dispersion curves of the pre-snapping configuration with bandgap from $196$-$242$Hz (black curves) and of the post-snap configuration with bandgap from $578$-$712$Hz (red curves).}
\label{fig:discr2}
\end{center}
\end{figure}

\newpage
The full nonlinear equations of motion of the two-mass resonator, excluding the aluminum host structure, were integrated in the time domain using a variable-step Runge--Kutta scheme (\texttt{ode45} in MATLAB). The latch state was kept constant within each integration segment and updated only at zero-crossing events. The system was simulated for $4$~s, with an initial excitation frequency $f_1$. At $t_{\mathrm{switch}}$, the base excitation frequency was switched from $f_1$ to $f_2$. The time-domain response shows that for $t < t_{\mathrm{switch}}$, the relative displacement remains bounded with $\max|\delta| \approx \SI{0.88}{mm}$, well below the engagement threshold $\delta_p=\SI{2.91}{mm}$. Consequently, the switching mechanism is not triggered. Immediately after the excitation frequency is changed to $f_2$, the out-of-phase resonance rapidly amplifies the relative displacement, causing $\delta$ to exceed $-\delta_p$ within approximately $\SI{4}{ms}$ and thereby engaging the latch. Following latching, the response settles into a sustained oscillation centered around $-\delta_{\mathrm{lock}}$. These dynamics are illustrated in Fig.~\ref{fig:discr2}a, which presents a spectrogram of the $x_2$ displacement. The signal $x_2(t)$ was resampled to $f_s \approx 8f_2$, ensuring adequate resolution above the Nyquist limit of the highest excitation frequency, and analyzed using a short-time Fourier transform (\texttt{spectrogram} in MATLAB). Since the time-frequency representation is computed directly from the nonlinear time-domain response rather than from a linearized frequency-response model, it captures both the resonant frequency selection and the harmonic content generated by the nonlinear coupling. The spectrogram confirms that the dominant response frequency follows the excitation frequency: the response is concentrated at $f_1$ prior to $t_{\mathrm{switch}}$ and transitions to $f_2$ after the latch is triggered. In addition, weaker components are visible at $2f_1$, $3f_1$, and $4f_1$ before the switch. These higher-order harmonics arise from the cubic spring nonlinearity present in the initial configuration.

Finally, dispersion curves are computed for the resonator attached to the host structure, using linearized models of both configurations.\ 
Specifically, the bistable element is represented by the stiffness $k_{2,1}$ in the unlatched configuration and by $k_{2,2}$ in the latched configuration.\ 
The resulting dispersion diagrams are shown in Figure~\ref{fig:discr2}b.\ 
Note that only the out-of-plane bending modes are visualized in the dispersion diagrams.\
For the unlatched configuration (pre-snap), a bandgap is observed around $200$~Hz ($196-252$~Hz), while a localized mode remains present near $600$~Hz.\ 
This mode corresponds to the second, out-of-phase eigenfrequency, for which $m_1$ exhibits only limited motion.\ 
After latching, the dispersion diagram shows a bandgap centered around $600$~Hz ($578-712$~Hz), demonstrating the ability of the proposed mechanism to substantially shift the bandgap location through a configuration change.\ 
For comparison, the unlatched configuration was redesigned with a mass ratio $m_1:m_2=1:1$, and the stiffnesses $k_1$ and $k_{2,1}$ were reselected to realize a conventional two-degree-of-freedom resonator with two fixed bandgaps.\ 
This design yields bandgaps of $200$-$227$~Hz and $595$-$653$~Hz.\
The comparison highlights that the proposed self-adaptive concept can achieve wider attenuation regions at both target frequencies by actively relocating the bandgap, rather than relying on multiple fixed bandgaps within a single configuration.

\section{Conclusion}
\label{sec:conclusion}
This work investigated two resonator concepts for locally resonant metamaterials that autonomously adapt their bandgap to the excitation frequency.\ 
The first concept employs a continuously moving slider mass, which passively adjusts the resonator eigenfrequency through self-positioning along a beam.\ 
While this mechanism successfully shifts the bandgap location, the slider tends to converge toward the central region of the beam, limiting the achievable tuning range and resulting in relatively narrow bandgaps.\ 
Next, a second concept based on vibration-induced bistable triggering was proposed.\ 
By switching between two discrete structural configurations, the resonator can relocate its bandgap from approximately $200$~Hz to $600$~Hz through a purely passive mechanism.\ Compared to a conventional fixed multi-bandgap design, the discrete concept provides wider attenuation regions at the targeted frequencies while maintaining a limited resonator mass.\ 
Overall, the results demonstrate the potential of passive self-adapting resonators for locally resonant metamaterials.


\section*{Acknowledgements}
The research of V.\ Cool (fellowship no.\ 1213925N) is funded by a grant from the Research Foundation - Flanders (FWO).\ 
The Research Fund KU Leuven is gratefully acknowledged for its support.\


\bibliography{References.bib}

\end{document}